%% file: authortemplate.tex
\documentclass[preprint2,twoside]{hwo}
\usepackage{enumitem}
\usepackage[table,dvipsnames]{xcolor}

\colorlet{LightBlue}{RoyalBlue!95!black!50}
\colorlet{MedBlue}{RoyalBlue!90!black!80}
\colorlet{DarkBlue}{RoyalBlue!50!black!100}

\input{hwo.h}

\begin{document}

\title{\textbf{\LARGE Resolving Individual Stars in Nearby Large Galaxies with the Habitable Worlds Observatory}}

\author {\textbf{\large Adam Smercina,$^{1,*}$ Tara Fetherolf,$^{2,\dagger}$}}
\affil{$^1$\small\it Space Telescope Science Institute, 3700 San Martin Dr., Baltimore, MD 21218, USA; $^*$NASA Hubble Fellow; {\rm \href{asmercina@stsci.edu}{asmercina@stsci.edu}}}
\affil{$^2$\small\it Department of Earth and Planetary Sciences, University of California Riverside, 900 University Avenue, Riverside, CA 92521, USA; $^\dagger$NASA Postdoctoral Program Fellow; {\rm \href{tara.fetherolf@gmail.com}{tara.fetherolf@gmail.com}}}

\author {\large \textbf{Contributing Authors:} Eric W. Koch,$^{3}$ Silvia Martocchia,$^{4,5}$ Chris Mihos,$^{6}$ Benjamin F. Williams$^{7}$}
\affil{$^{3}$ National Radio Astronomy Observatory, 800 Bradbury SE, Suite 235, Albuquerque, NM 87106}
\affil{$^{4}$ Aix Marseille Universit\'{e}, CNRS, CNES, LAM, Marseille, France}
\affil{$^5$ Astronomisches Rechen-Institut, Zentrum f\"{u}r Astronomie der Universit\"{a}t Heidelberg, M\"{o}nchhofstra{\ss}e 12-14, D-69120 Heidelberg, Germany}
\affil{$^{6}$ Department of Astronomy, Case Western Reserve University, 10900 Euclid Ave., Cleveland, OH 44106, USA}
\affil{$^{7}$ Department of Astronomy, Box 351580, University of Washington, Seattle, WA 98195, USA}

\author {\small \textbf{Endorsed by:} Gagandeep Anand (STScI), Borja Anguiano (CEFCA), Pauline Barmby (Western University), Breanna Binder (Cal Poly Pomona), Howard Bond (Penn State University), Lisa Bugnet (ISTA), Luca Casagrande (Australian National University), Christopher Clark (STScI), Ruben Joaquin Diaz (NOIRLab), Meredith Durbin (UC Berkeley), Helena Faustino Vieira (Stockholm University), Carl Grillmair (California Institute of Technology), Marziye Jafariyazani (SETI Institute / NASA Ames Research Center), Brad Koplitz (Arizona State University), Eunjeong Lee (EisKosmos (CROASAEN), Inc.), Erik Monson (Penn State University), Ilaria Musella (INAF Osservatorio Astronomico di Capodimonte), Donatas Narbutis (Institute of Theoretical Physics and Astronomy, Vilnius University), Gijs Nelemans (Radboud University), Anna O'Grady (Carnegie Mellon University), Iain Neill Reid (STScI), Nicol\'{a}s Rodr\'{i}guez-Segovia (UNSW Canberra), Frank Soboczenski (University of York, King's College London), Heloise Stevance (University of Oxford), Grace Telford (Princeton University), David Thilker (Johns Hopkins University)}



\begin{abstract}
  The varied and dynamic evolutionary histories of galaxies give rise to the stunning diversity in their properties that we observe in the present-day universe. HST, and now JWST, have pioneered the study of resolved individual stars in the Milky Way and other members of the Local Group, uncovering the drivers of their morphological, star formation, and chemical evolution. HWO will constitute a paradigm shift: introducing the ability to panchromatically resolve the main bodies of every galaxy in the Local Volume into their constituent stars. In this science case, we summarize the breakthrough progress that HWO will advance in the field of galaxy evolution through resolved stellar populations. HWO will transform our understanding of galaxies in three distance regimes: (1) in the nearest galaxies ($\sim$5\,Mpc), where it will resolve stars below the oldest Main Sequence Turnoff, enabling precision stellar astrophysics and star formation history (SFH) inferences to the earliest cosmic times; (2) in the greater Local Volume ($\sim$20\,Mpc), where it will resolve stars below the Red Clump, providing access to accurate SFHs for hundreds of galaxies, spanning the entire Hubble Sequence; and (3) out to cosmological volumes ($\sim$50+\,Mpc), providing access to the luminous stellar populations in thousands of galaxies, enabling unprecedented views of their morphology, stellar abundances, and dust content. The principal technological requirement advanced by this science case is a camera with a resolution of $\leqslant$0\farcs015 that is diffraction-limited, and Nyquist-sampled (0\farcs01 per pixel), to at least 550\,nm -- comparable to the High Definition Imager from the LUVOIR concept. 
  \\
  \\
\end{abstract}

\vspace{2cm}

\section{Science Goal}
\vspace{-5pt}
\noindent{\textsc{How do other galaxies evolve similarly or differently from the Milky Way?}} \\

Galaxies are incredibly diverse objects; they display an enormous range of different shapes, sizes, and compositions. Much of this diversity can likely be explained by the evolutionary trajectory each galaxy has taken. We know that our own Milky Way, for example, has led an active social life, interacting with many other galaxies over the past 13 billion years. We have also begun to connect properties of the Milky Way, and the stars that live in it, to the details of these social interactions. However, the Milky Way is one galaxy; how have the different social lives of other galaxies in the known universe resulted in differences, or similarities, between them and our Galaxy? And what are the key drivers of the development of morphological structures, such as spiral arms, which are not present during earlier cosmic times?

In order to access these galaxy archaeological histories, we need to access the information encoded in their longest-lived visible components -- their stars. These individual stars are exceptionally faint at the distances of external galaxies, and require powerful telescopes with sharp imaging to detect them. HWO has the potential to surpass the capability of both HST and JWST in allowing the archaeological study of external galaxies in the nearby universe at a similar level of detail to our current understanding of the Milky Way -- an enormous stride forward in the study of galaxy evolution. We will detail HWO’s potential to contribute to our understanding of galaxy evolution through resolved stars in three distance regimes: (1) resolving individual stars in external galaxies, out to 5 Mpc, with equivalent fidelity to HST’s and JWST’s current capabilities in the Local Group; (2) reconstructing resolved star formation and chemical enrichment histories in galaxies out to 20 Mpc -- increasing the accessible number of galaxies by two orders of magnitude; and (3)  first-ever access to the structure of galaxies traced by their luminous individual stars out to the edges of the ``Hubble Flow'' (out to 50 Mpc). 

This science is directly related to multiple key science areas identified by the Astro2020 decadal review, Pathways to Discovery in Astronomy and Astrophysics for the 2020s \citep{NationalAcademiesofSciences2021}, including Unveiling the Hidden Drivers of Galaxy Growth, Other Worlds and Suns in Context, New Windows on the Dynamic Universe, and Cosmic Ecosystems. Among the prime examples of potential breakthrough discoveries, the Astro2020 review states: 

\newenvironment{myquote}{
  \begin{itemize}[topsep=-5pt,rightmargin=3pt]
  \item[]}{\end{itemize}}

\begin{myquote}
{\it This telescope will be capable of achieving breakthrough discoveries across nearly all of astrophysics. Prime examples include\ldots the construction of stellar fossil histories of the galaxies in the neighborhood of the Milky Way. These examples all constitute major components on the New Windows on the Dynamic Universe and the Unveiling the Hidden Drivers of Galaxy Growth priority science areas, and they represent only the tip of the iceberg of the impact such a telescope would have.''} -- Chapter 7, Page 200 \\
\end{myquote}

By using resolved individual stars to reconstruct the star formation and chemical enrichment histories of galaxies throughout the Local Volume, HWO will revolutionize our understanding of the detailed evolution of galaxies, the fundamental demographics of stars, and their global influence on the interstellar medium. This unprecedentedly detailed view of hundreds of diverse galaxies requires the precise measurement of panchromatic photometry of individual stars in crowded regions to deep limits. This requires a unique blend of resolution, wavelength coverage, FOV, and image stability. No ground-based instrument -- not even the upcoming suite of Extremely Large Telescope (ELT) facilities -- can provide a paradigm shift in this science comparable to HWO. By unlocking at cosmic volumes what not long ago was strictly Milky Way science, HWO will truly be key to understanding the drivers of galaxy growth. 

\section{Science Objective}
\label{sec:objectives}

HST has defined the study of deep resolved stellar populations in the nearby universe. From the star formation histories (SFHs) of globular clusters in the Milky Way, to those of dwarf galaxies in the Local Group and near-field, the resolution and sensitivity gained from space-based imaging has been critical to resolved stellar population studies for 35 years, and JWST has now taken this torch. The primary hurdles to progress for this field are two-fold: 
\begin{enumerate}[topsep=3pt,leftmargin=12pt,itemsep=0pt]
    \item Depth -- stars are faint, and resolving them at high-S/N requires sensitive cameras on powerful telescopes; and 
    \item Resolution -- stars in galaxies are close together, and individually measuring them requires high-resolution imaging with a well-characterized point-spread function. When stars blend together, this is known as ``crowding'' and it fundamentally limits the depth of the measurable photometry. 
\end{enumerate}
In the following three objectives (\S\,\ref{sec:depth1}--\ref{sec:depth3}), we summarize the progress that HWO can achieve in this field over JWST and HST. For a ~6m space-telescope, with modern detectors and cameras, operating in the optical, we find that the key technical specification needed to achieve these gains in resolved stellar populations is sampling of the PSF. As laid out by the Large UV/Optical/Infrared Surveyor mission report \citep{TheLUVOIRTeam2019}, the cameras must be well-sampled and diffraction-limited in the optical to take advantage of HWO’s theoretical resolution increase, and make substantial progress in resolved star science. 

\begin{figure}[t]
    \centering
    \includegraphics[width=\linewidth]{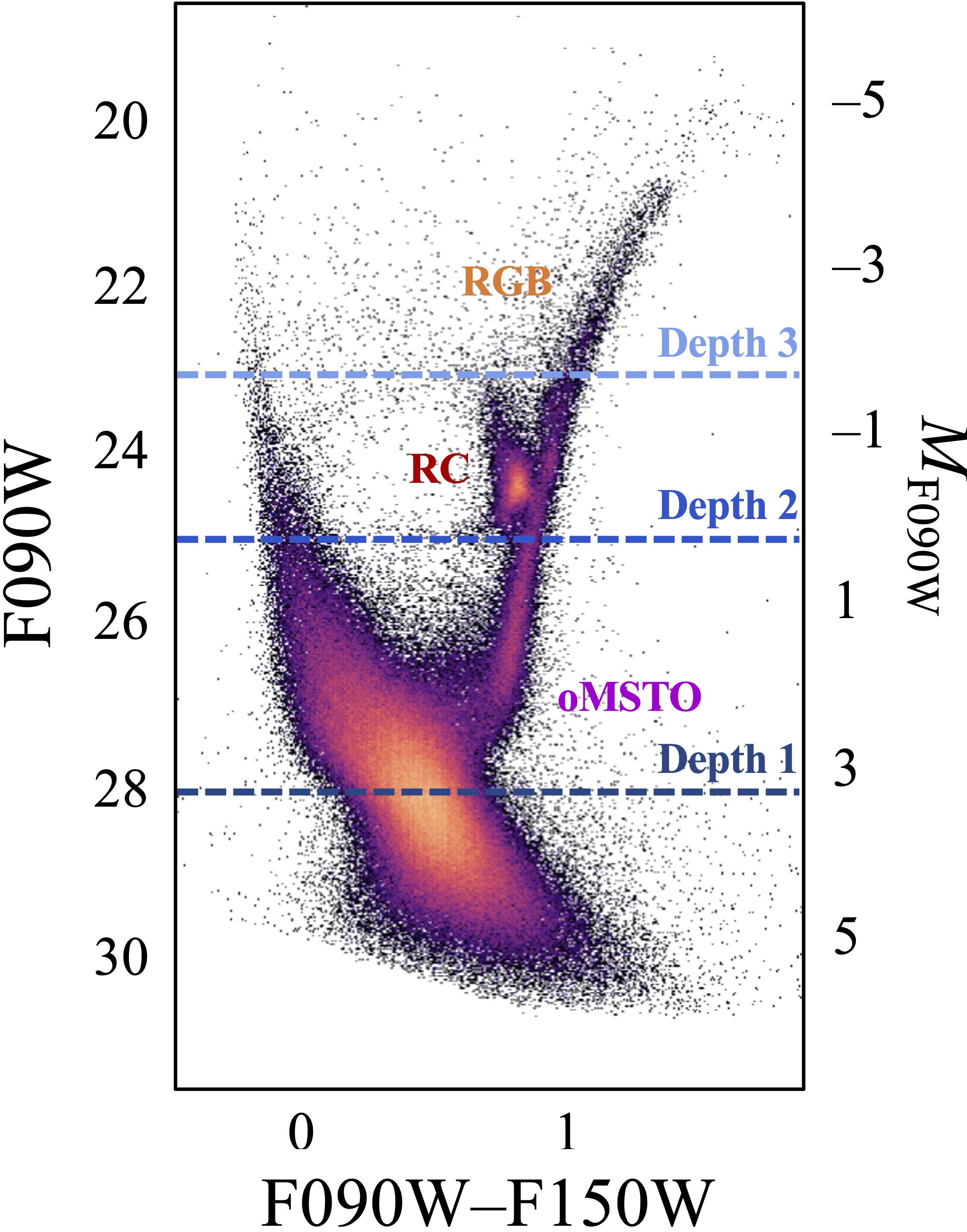}
    \caption{An example color-magnitude diagram (CMD) of stars detected in the galaxy WLM with JWST NIRCam observations, obtained by the JWST Resolved Stellar Populations ERS team \citep{McQuinn2024,Weisz2024}. CMD adapted from Figure 7 of \cite{Weisz2024}. The \textbf{3 Depth Levels} that provide access to different aspects of the stellar population are labeled, corresponding to 1 mag below the oldest Main Sequence Turnoff (oMSTO) and Red Clump (RC), and 2 magnitudes below the Tip of the Red Giant Branch (RGB).}
    \label{fig:cmd-depth}
\end{figure}

The information that can be extracted from stars is dependent on depth, which is dependent on both resolution and sensitivity in a highly non-linear way. We organize the most impactful science into three sections based on relative depth, i.e. absolute magnitude. We summarize these three depth levels in Fig. \ref{fig:cmd-depth}. 

\textcolor{LightBlue}{\subsection{Resolving individual stars, including individual stellar clusters, in the nearest large galaxies with Local Group-like fidelity.}}
\label{sec:depth1}

While many techniques have been developed to infer the star formation histories of star clusters and galaxies, comparing photometry of individual stars to depths 1 magnitude below the oldest main sequence turn-off (oMSTO) (resolving the feature at S/N=10) to models of stellar evolution is the most accurate method. For complex populations, such as those found in galaxies, this requires resolving stars with ages of 12-13 Gyr, which are very faint. HST has provided access to deep photometry, and accurate SFHs, for the nearest star clusters and dwarf galaxies \citep[][and many others]{Stetson1994,Dolphin2002,Weisz2014}, and JWST is improving on HST’s legacy by pushing out to comparable relative depths ($M_I\,{\sim}\,3$) in the near-field \citep[e.g.,][]{McQuinn2024}. JWST is improving on HST not just in sensitivity, but also in resolution, which increases depth in crowded stellar fields. However, neither HST nor JWST can reach the oMSTO for stellar populations beyond the near-field ($<$2--2.5\,Mpc), which is insufficient to reach the nearest groups of galaxies comparable to the MW and M31. 

\begin{figure}[t]
    \centering
    \includegraphics[width=0.99\linewidth]{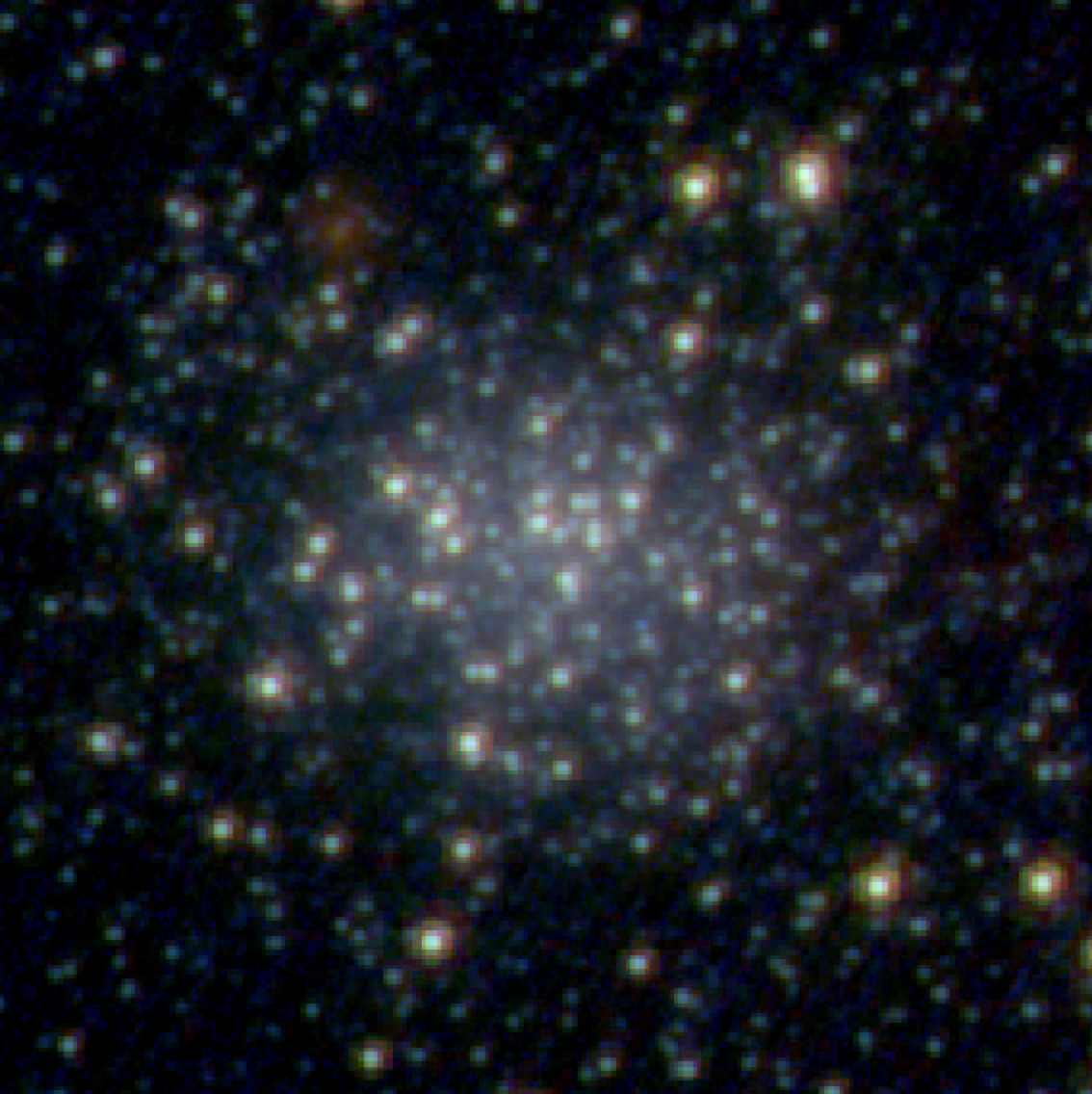}
    \caption{A 7\arcsec$\times$7\arcsec\ ($\sim$28.5 pc\,$\times$\,28.5 pc) James Webb Space Telescope NIRCam color image (F090W / F200W / F335M) of a resolved young stellar cluster in the Local Group galaxy Messier 33. Young clusters such as this are only currently resolvable in galaxies out to the near-field, 2--2.5 Mpc. HWO will stretch this to at least 5 Mpc.}
    \label{fig:star-cluster}
\end{figure}

In addition to the information that can be gleaned from individual stars regarding the evolutionary histories of galaxies, young star clusters ($<$100 Myr) are tracers of the fundamental process of star formation in galaxies \citep[e.g.,][]{Krumholz2014}. Young clusters are dynamically unevolved and therefore serve as direct tracers of the stellar demographics resulting from a single star-forming event. As such, these objects provide direct measurement of the high-mass end of the stellar IMF if they can be resolved into their individual stellar populations (see Fig. \ref{fig:star-cluster}). Studies with HST have resolved young star clusters in star-forming Local Group galaxies such as M31 \citep{Weisz2015} and M33 \citep{Wainer2024}. However, even with the increase in resolution from JWST, these clusters are only currently resolvable in galaxies out to the near-field, 2--2.5 Mpc -- which comprises a very limited set of star-forming galaxies and does not fully trace how the IMF may change with metallicity and star formation intensity.  

\begin{figure*}[t]
    \centering
    \includegraphics[width=\linewidth]{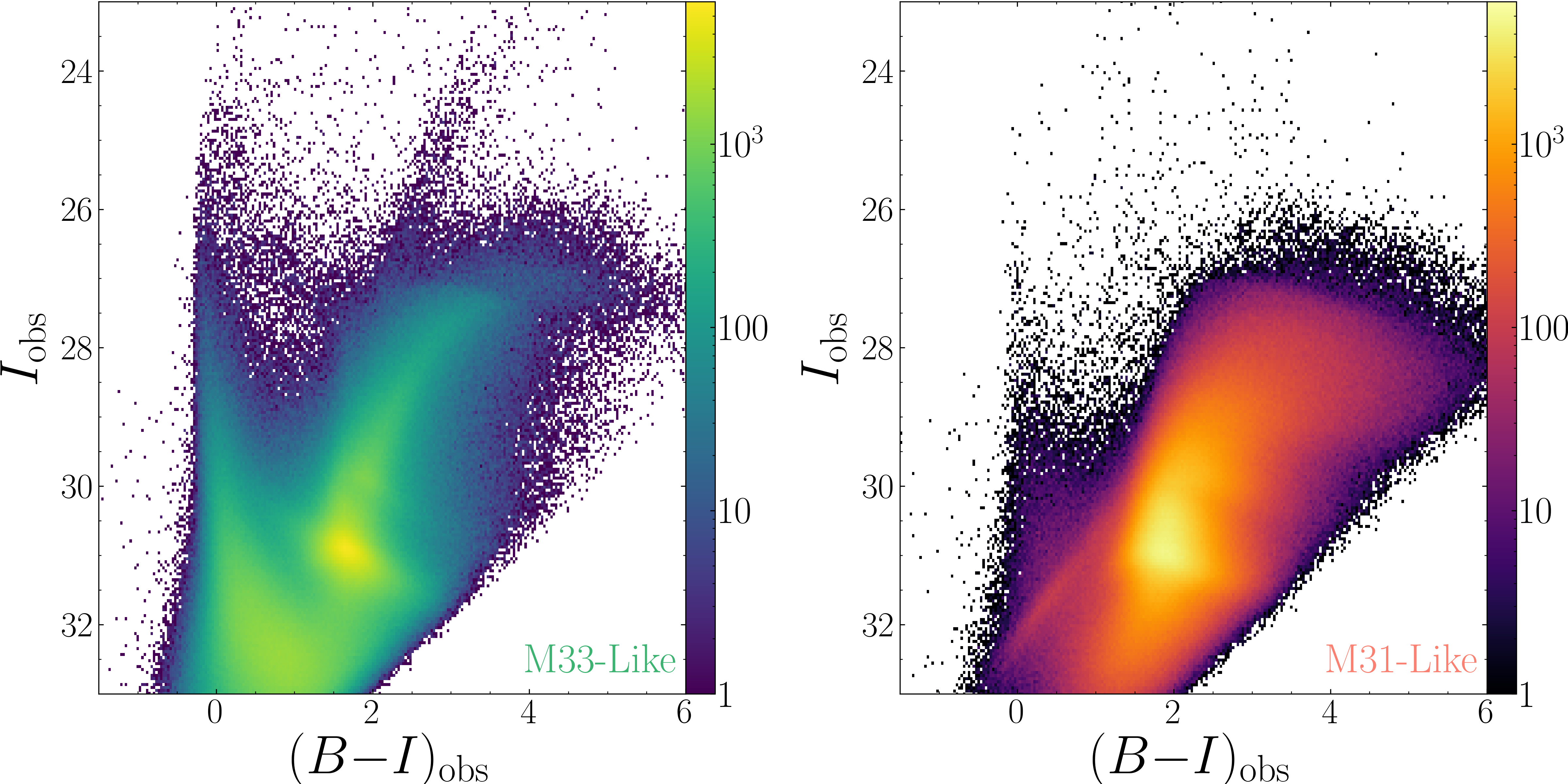}
    \caption{Representative color--magnitude diagrams (CMDs) of resolved stars detected in the Local Group galaxies M33 (left) and M31 (right), but placed at a distance of 16 Mpc. For stellar surface densities similar to those found in large disk galaxies, like M33, and for the camera design proposed in the LUVOIR report, these are examples of CMDs that could be resolvable in galaxies at 12-20 Mpc with HWO. At these depths of $>$31 magnitudes in I-band, features such as the Red Clump and blue Horizontal Branch are resolved, making the ancient star formation history of the population accessible using modern CMD fitting techniques. CMDs reproduced from the PHATTER \citep{Williams2021} and PHAT Legacy \citep{Williams2023} stellar catalogs.}
    \label{fig:m33-m31}
\end{figure*}

\noindent\textbf{\underline{HWO’s Advantage}:} With its substantially higher-resolution, sensitivity, and PSF sampling, HWO will provide brand-new access to accurate and precise SFHs, and resolved IMF measurements, in every star-forming galaxy within 5\,Mpc ($\sim$100 galaxies) with stellar mass $10^8{-}10^{11}\,M_{\odot}$, comprising a broad range in metallicity and star formation properties. There are currently only 4 external galaxies accessible with this precision via HST and JWST: M31, M33, and the Magellanic Clouds. Photometry to relative depths equivalent to JWST will be possible with HWO out to 5 Mpc, twice as far as JWST, providing access to the complete SFHs of all of the nearest large galaxies, as well as much more numerous lower-mass dwarf galaxies. Young stellar clusters would also be resolvable with HWO out to 5 Mpc, a factor of 3-5 compared to JWST, allowing an independent measure of the stellar IMF in thousands of clusters across dozens of galaxies with diverse properties. 

\textcolor{MedBlue}{\subsection{Panchromatic imaging and photometry of individual resolved stars in large galaxies throughout the Local Volume to deep limits.}}
\label{sec:depth2}

While accessing the oMSTO, and therefore the complete lifetime SFHs of galaxies will be challenging for HWO beyond 5 Mpc -- due to both the faintness of the this feature and the associated stellar crowding -- it will be able to access intermediate features of post-main sequence low-mass stellar evolution, such as the Red Clump (RC), to much greater distances, and importantly, in more massive galaxies. Similar to the oMSTO, the RC is an important evolutionary pileup point in stellar CMDs, which can break the age--metallicity--reddening degeneracy within certain ranges of the age--metallicity parameter space. Generally the RC retains age sensitivity for SFH fitting to ages of 6\,Gyr \citep[][and others]{Weisz2014,Williams2017}. Taking advantage of the RC for SFH fitting requires detecting stars to depths 1 magnitude below the feature ($M_I\,{\sim}\,0$), corresponding to a S/N = 10 for the feature itself. 

\begin{figure*}[ht!]
    \centering
    \includegraphics[width=0.97\linewidth]{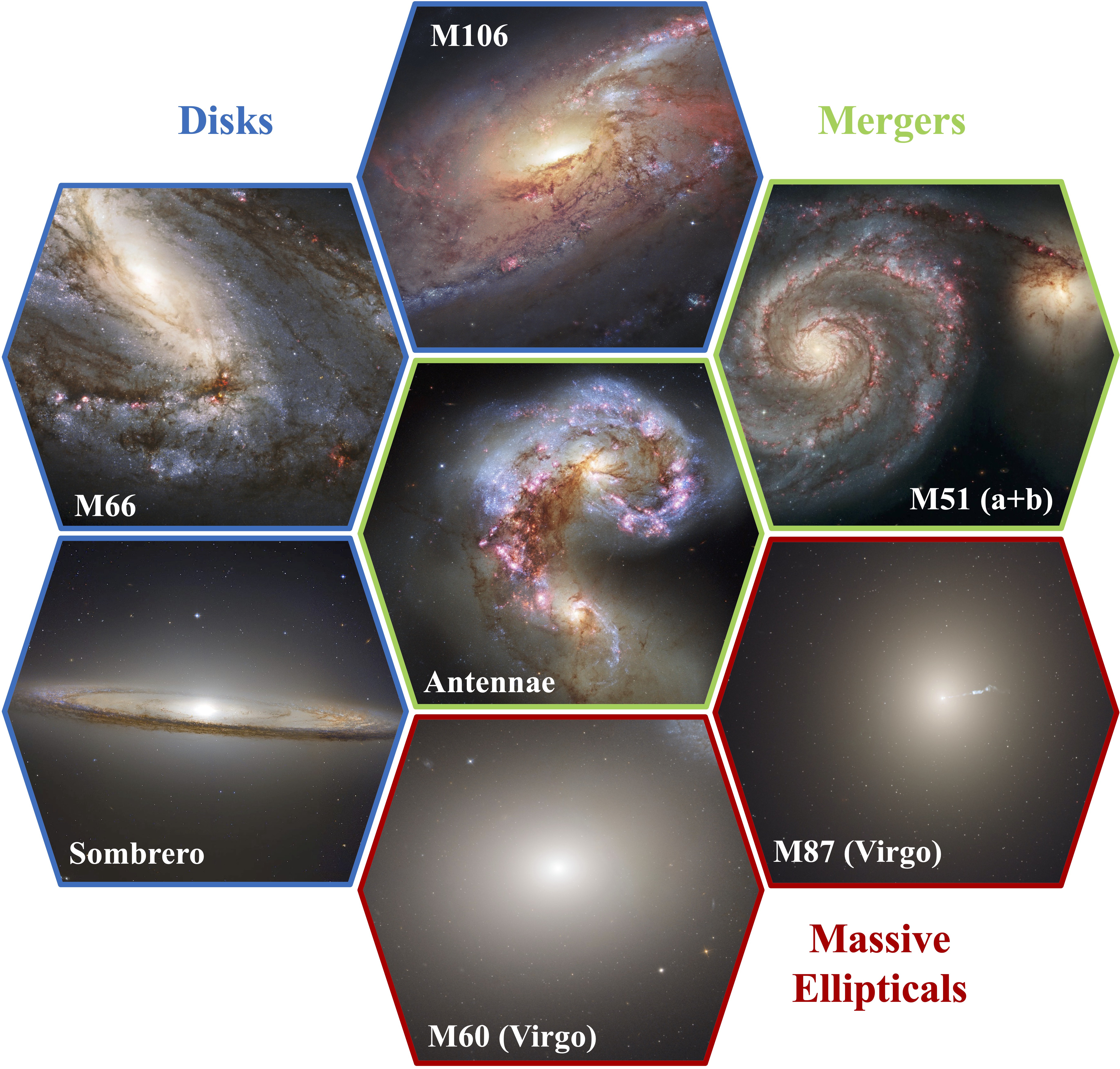}
    \caption{HST images of well-known examples of nearby galaxies showcasing the diversity of galaxy evolutionary histories that could be probed with HWO at distances $\sim$5--20\,Mpc, greatly expanding on the limited galaxy populations of the local universe. Late-stage mergers (e.g., M51 and the Antennae), disrupted disks with active galactic nuclei (e.g., M66, M106, the Sombrero), and massive ellipticals (e.g., M87 and M60 in the Virgo Cluster) would all be within HWO’s reach. While HST provided the first detailed pictures of these galaxies at high-resolution, HWO will provide a leap forward --- it will be capable of turning images such as these into star fields, resolving their constituent stars to below the Red Clump and revealing their evolutionary histories in detail. \vspace{3pt}\\
    \textit{Image Credit}: Antennae(NGC 4038/4039)---HST $U/B/V/R/I/H\alpha$\ mosaic (NASA, ESA, and the Hubble Heritage Team); M106---$G/V/R/I/H\alpha$\ combined mosaic from HST and ground-based observations (NASA, ESA, the Hubble Heritage Team, R. Gendler, J. GaBany, and A. Riess); M66---HST $V/I/H\alpha$\ (NASA, ESA, the Hubble Heritage Team, Davide De Martin, and Robert Gendler); M51---HST $B/V/I/H\alpha$\ mosaic (NASA, ESA, S. Beckwith, and the Hubble Heritage Team); Sombrero(M104)---HST $B/V/r$\ mosaic (NASA and the Hubble Heritage Team); M87---HST $g/V/I$\ mosaic (NASA, ESA, the Hubble Heritage Team, P. Cote, E. Baltz); M60---HST $B/V/I$\ mosaic (NASA, ESA, and A. Seth).
    }
    \label{fig:example-galaxies}
\end{figure*}

The limit of HST’s reach to measure the RC feature is $\sim$3.5 Mpc, which has been achieved for only a single field in a single massive galaxy outside of the Local Group, M81 \citep{Williams2009}, as well as the stellar halos of M81 and Centaurus A \citep{Rejkuba2005,Durrell2010}. Reaching the required depth of $\sim$29 magnitudes in $I$-band is incredibly time-intensive with HST. In the past decade, HST has helped to reveal the formation and evolutionary history of nearby galaxies in the Local Group by mapping the nearby large disk galaxies M31 \citep[e.g.,][]{Dalcanton2012,Williams2015} and M33 \citep[e.g.,][]{Williams2021,Smercina2023} to RC depths (Fig. \ref{fig:m33-m31}). These surveys have demonstrated the breadth of scientific insight that can be gained by mapping the global stellar populations of large galaxies to below the horizontal branch, often generating tens-to-hundreds of millions of stars. Efforts using JWST will further this science by mapping stars to RC depths in the nearest galaxy groups, such as M82 (GO-5145, Cycle 3), and will likely permit surveys of this type out to 5\,Mpc.

\noindent\textbf{\underline{HWO’s Advantage}:} With its superior resolution and sensitivity, HWO is capable of providing accurate resolved SFHs (via CMD fitting) for every galaxy out to 20\,Mpc by resolving stars to 1 magnitude below the Red Clump (S/N @ RC = 10). In comparison, JWST is limited (largely by resolution) to $\sim$5\,Mpc for similar measurements. Consequently, there is not a single late-stage merger, massive elliptical, or cluster environment accessible with JWST. Panchromatic imaging and photometry of individual resolved stars in large galaxies throughout the Local Volume to deep limits would facilitate spatially-resolved CMD fitting and fitting of individual stellar SEDs, which can be used to calculate global star formation and metal enrichment histories. Such surveys would be comparable to the Panchromatic Hubble Andromeda Treasury (PHAT; \citealt{Dalcanton2012}), but for galaxies 20 times more distant. The PSF size and sampling is key here: a 6m mirror with an imager designed similarly to LUVOIR’s HDI instrument \citep{TheLUVOIRTeam2019} could permit crowded field photometry (UV/Opt) out to $\sim$20 Mpc. Furthermore, the study of these galaxies would be far more efficient than surveys such as PHAT, as a larger fraction of the galaxies would be visible in a single HWO FOV. This would expand the study of detailed galaxy evolutionary histories, and related stellar evolutionary science, to a much broader range of objects (see Fig. \ref{fig:example-galaxies}), including galaxies like M51, M106, the Sombrero, the Antennae, and even M87 and the Virgo Cluster. 

\begin{figure*}[t]
    \centering
    \begin{minipage}{0.5\linewidth}
        \includegraphics[width=\linewidth]{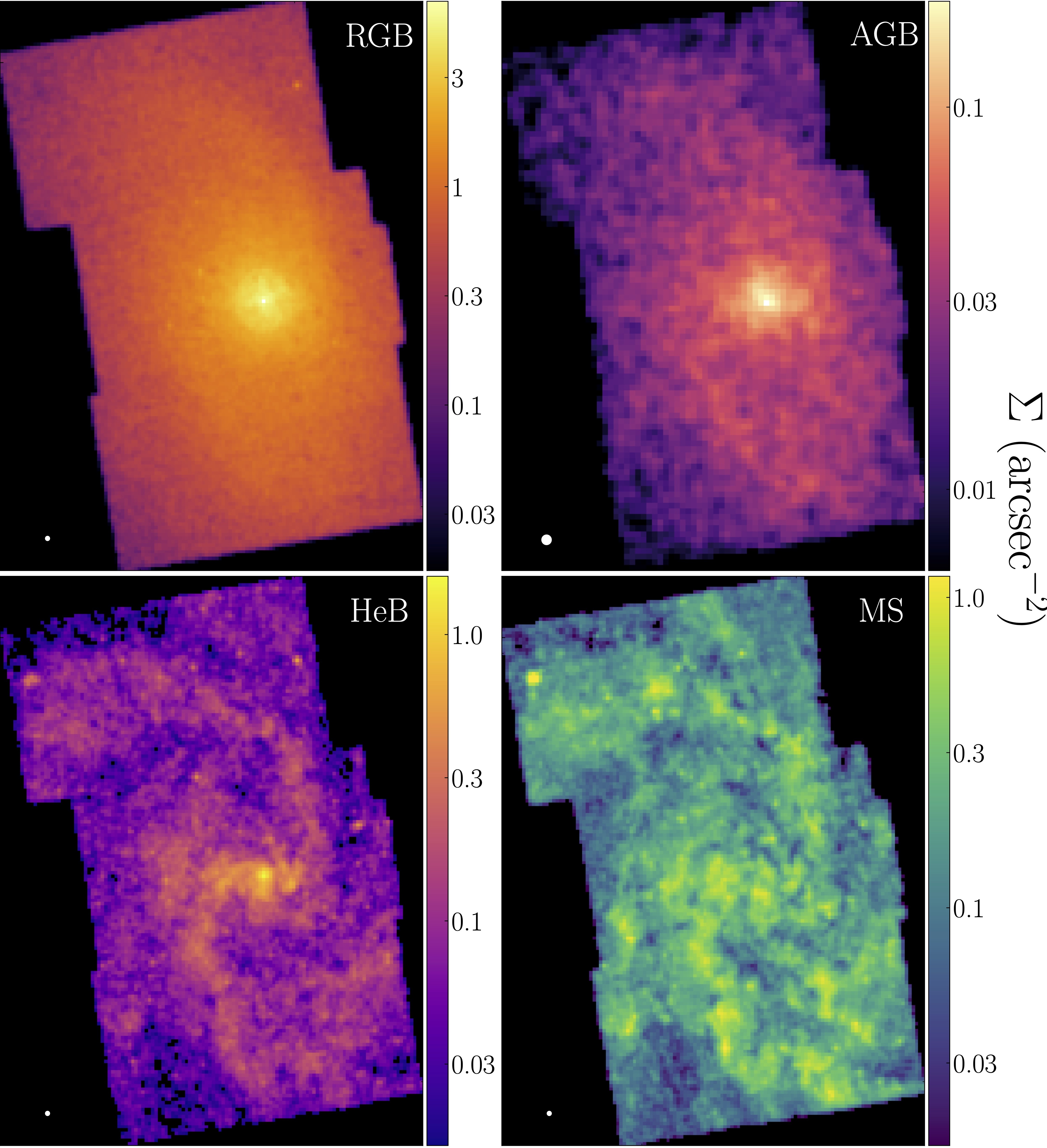}
    \end{minipage}
    \hspace{5pt}
    \begin{minipage}{0.475\linewidth}
        \includegraphics[width=\linewidth]{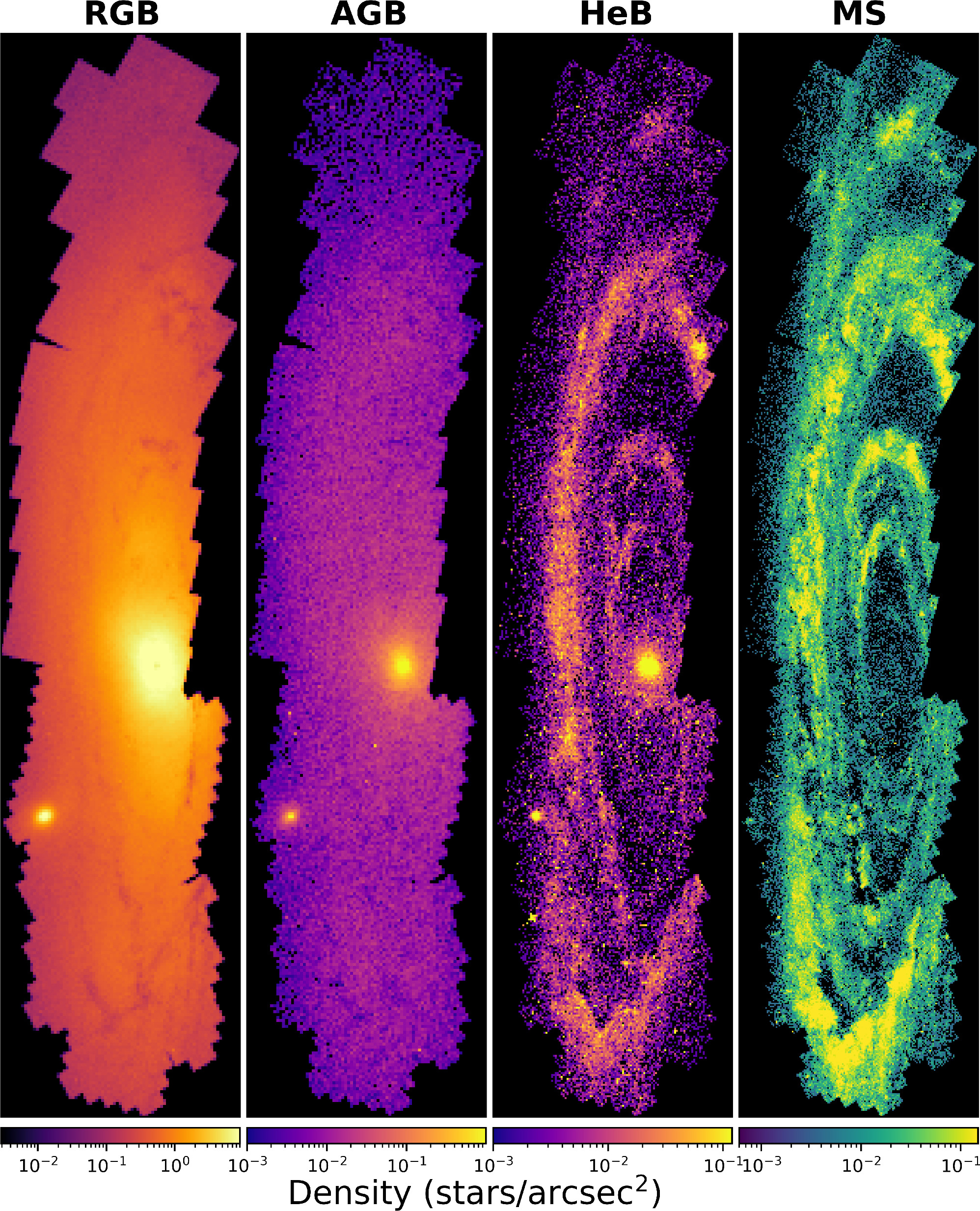}
    \end{minipage}
    \caption{Maps of surface density of resolved individual upper RGB (old, evolved), upper AGB (intermediate, evolved), Core Helium-burning (HeB; intermediate-young), and upper Main Sequence (young, massive; right) stars in M33 (left; PHATTER Survey, \citealt{Smercina2023}) and M31 (right; PHAST Survey, \citealt{Chen2025}). Constructing these maps required considerable mosaicing with HST, as both galaxies are large on the sky. As more distant galaxies have much smaller angular sizes, a large increase in resolution and a similar imaging FOV would allow HWO to map hundreds of galaxies with this same level of detail much more efficiently.}
    \label{fig:maps}
\end{figure*}

Beyond the ability to reach the RC in the disks of galaxies, HWO will also provide unparalleled access to both the innermost and outermost regions of many galaxies. In the densest parts of galaxies, the pileup of stars at the RC causes a rapid increase in stellar density once the feature is reached, making it particularly susceptible to crowding. Even for the nearest large galaxies, such as M31, M32, and M33, HST struggles to reach the depth of the RC in the most crowded regions of galaxies, such as their nuclear or bulge regions \citep[e.g.,][]{Grillmair1996,Monachesi2011,Williams2017,Lazzarini2022}. The stellar populations occupying the central regions of elliptical galaxies and spiral bulges are particularly enigmatic. Despite a lack of ongoing star formation, dense stellar regions, from the local compact elliptical M32 and the bulge of M31 to distant ellipticals, show a mysterious ``UV upturn'' \citep[e.g.,][]{Code1979,Greggio1990,Brown1997,O'Connell1999}. This phenomenon likely constrains a critical aspect of stellar evolution on the Horizontal Branch \citep{Brown2004,Brown2008}. It is a fundamental, and unsolved, limitation on our ability to infer SFHs. With its unprecedented UV resolution and sensitivity, HWO is perfectly poised to solve this long-standing problems. 

Stellar halos, meanwhile, are the least dense regions of galaxies. They encode the accreted stellar debris from past merger events, as shown in the Milky Way and a number of the nearest large galaxies, including panoramic examples using resolved into stars (M31, \citealt{martin2013,Ibata2014}; M81, \citealt{okamoto2015,smercina2020}; Centaurus A, \citealt{Crnojevic2016}; and M64, \citealt{Smercina2023b}). While the Roman Space Telescope will soon bring similar panoramic views of the luminous stellar halo populations in nearby galaxies \citep[e.g.,][]{Akeson2019,Lancaster2022}, HWO will be able to reach the oMSTO (as in \S\,\ref{sec:depth1} for the main bodies of much closer galaxies) --- allowing precision age-dating of the accreted stellar debris, and therefore the timescale of past mergers \citep{Brown2006}. This would be particularly revolutionary in clusters of galaxies such as Virgo, where the intracluster light traces the assembly of these cosmologically-important structures \citep[e.g.,][]{Mihos2017}.

\textcolor{DarkBlue}{\subsection{Resolving Bulk Luminous Stellar Populations to Reveal Galaxy Structure throughout the Low Redshift Universe}}
\label{sec:depth3}

\begin{figure}[t]
    \centering
    \includegraphics[width=\linewidth]{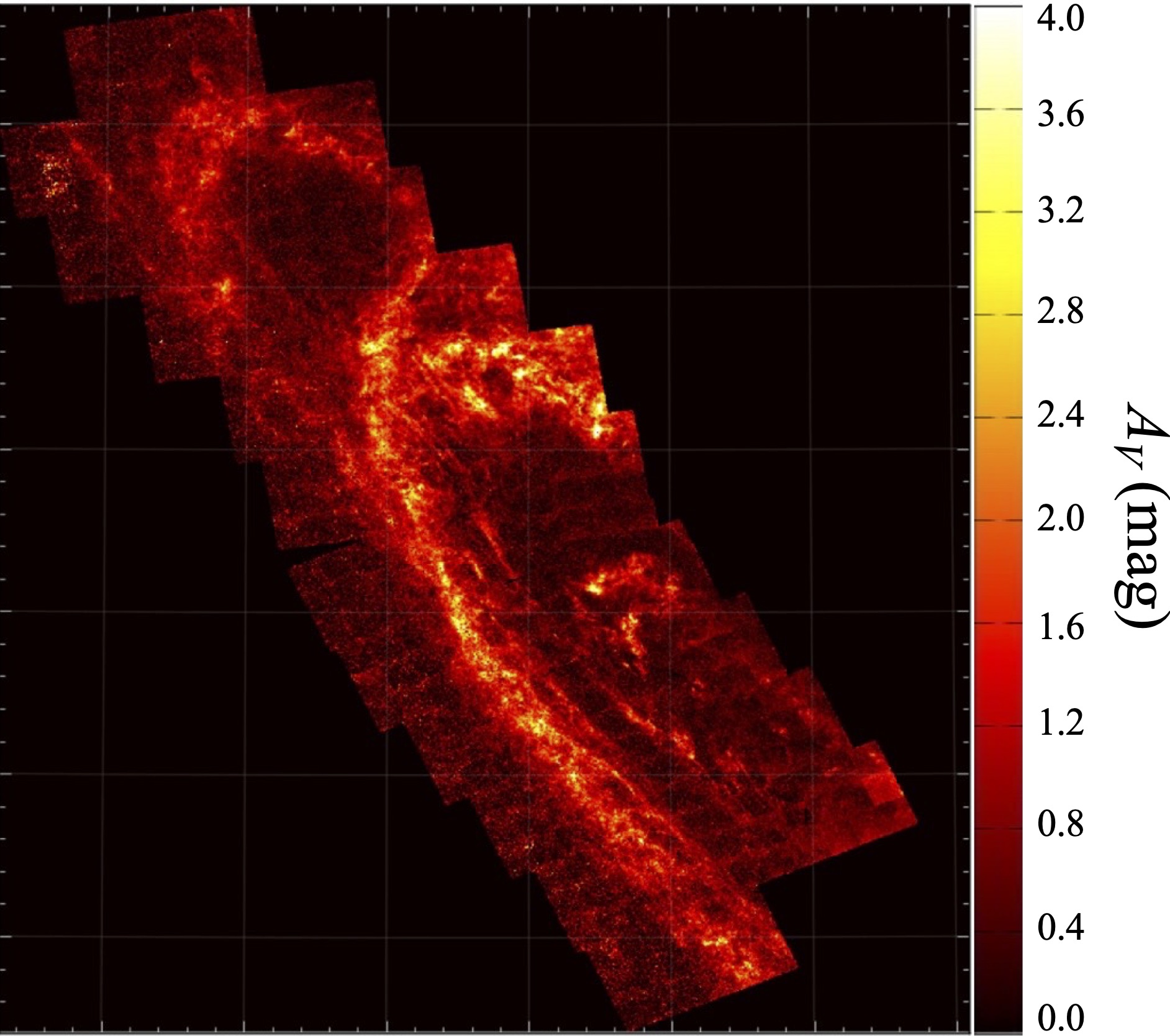}
    \caption{High-resolution map of line-of-sight extinction ($A_V$) in M31 from PHAT, measured from the reddening of individual luminous RGB stars \citep{Dalcanton2015}. Due to their brightness, luminous populations will be detectable and uncrowded with HWO in galaxies out to $\sim$50\,Mpc, marking the first time that the structure and extinction properties of galaxies at cosmological distances will be able to be studied using individual stellar populations.}
    \label{fig:dust-map}
\end{figure}

While stellar densities will likely be too high to resolve stars to RC depths past 20 Mpc, the incredible sensitivity and high-resolution of HWO will enable the study of luminous individual stars (Depth Level \#3; Fig. \ref{fig:cmd-depth}) to much greater distances -- including into the low-redshift universe. Resolving bright Main Sequence, core Helium Burning (HeB), Red Giant Branch (RGB), and Asymptotic Giant Branch (AGB) stars will permit measurements of the geometric distribution of individual stars, and their relation to galactic-scale properties, such as  galaxy structures (spiral arms, bars, etc.; e.g., \citealt{Smercina2023,Williams2023}), recent star formation \citep[e.g.,][]{Lazzarini2022}, and dust \citep[e.g.,][]{Dalcanton2015,Dalcanton2023,Lindberg2025}. Such maps of stellar density and dust extinction have only been constructed in the nearest galaxies with HST (see Figs.\,\ref{fig:maps} \&\ \ref{fig:dust-map}), and have yet to be constructed with JWST. While JWST will be limited to galaxies within 10 Mpc, and only in the near-infrared, HWO will provide panchromatic UV--IR imaging for these luminous populations out to \textit{at least 50\,Mpc}. 

Pairing stellar maps, like those shown in Fig. \ref{fig:maps}, with dynamical properties from high-resolution IFU spectroscopy with HWO would allow chemo-dynamical analysis in many of these thousands of galaxies \citep[e.g.,][]{Gibson2024}. Furthermore, the resolved distributions of these stars in the main bodies and outskirts of galaxies will reveal traces of previous merging events, as has been shown in many nearby galaxies (e.g., \citealt{Ibata2014,Crnojevic2016,Smercina2023b}; see \S\,\ref{sec:depth2}).

Moreover, the panchromatic nature of HWO would allow individual extinction values for the millions of stars detected across the faces of these galaxies \citep[e.g.,][]{Lindberg2025}, which would in turn pave the way for true 3D dust mapping for the entire Local Volume (and beyond). Pairing 3D dust mapping with existing maps of the ISM in these galaxies \citep[e.g., PHANGS;][]{Thilker2023,Williams2024} would provide critical constraints on the vertical structure of the neutral ISM. Such 3D views of the Milky Way are only just now capable of testing modern star formation theories \citep[e.g.,][]{Ostriker2011} in the Solar Neighborhood \citep[e.g.,][]{Alves2020}. HWO's global views of galaxies' stars will therefore enable enormous strides in our fundamental understanding of star formation. 

\noindent\textbf{\underline{HWO’s Advantage}:} As has been seen in the HST surveys of M31 and M33 in the Local Group, the structures of galaxies, their active star formation, and dust properties can differ substantially from the inferences drawn from integrated light measurements. JWST’s theoretical reach for spatial mapping of these luminous resolved stars extends to $\sim$10 Mpc -- encompassing $\sim$30 MW-mass galaxies, and $\sim$50 dwarf spirals comparable in mass to the LMC/M33. HWO will provide these measures for every large galaxy out to 50\,Mpc, increasing these to $\sim$3000 and $\sim$1500, respectively -- \textit{a two orders of magnitude increase in accessible systems}. This is comparable to the increase in stellar demographics of the Milky Way with the advent of Gaia, or in exoplanet demographics with the launch of Kepler. \\

\subsection{Summary}
\label{sec:obj-summary}

Overall, a high-resolution, high-sensitivity, and panchromatic imaging instrument on HWO will enable a paradigm shift in our understanding of star formation, chemical enrichment, and structural evolution in galaxies. It would allow us, for the first time, to measure complete SFHs and the high-mass IMF in a diverse set of nearby massive galaxies; measure the intermediate SFHs and chemical enrichment histories for hundreds of massive galaxies out to 20\,Mpc, including in previously untouched environments such as the Virgo Cluster; and map the structure and recent star formation of galaxies out to 50\,Mpc using individual stars. In short, HWO could provide the most significant improvement in our understanding of galaxy evolution in a generation.

\section{Physical Parameters}
\label{sec:params}

\begin{figure*}[t]
    \centering
    \includegraphics[width=0.95\linewidth]{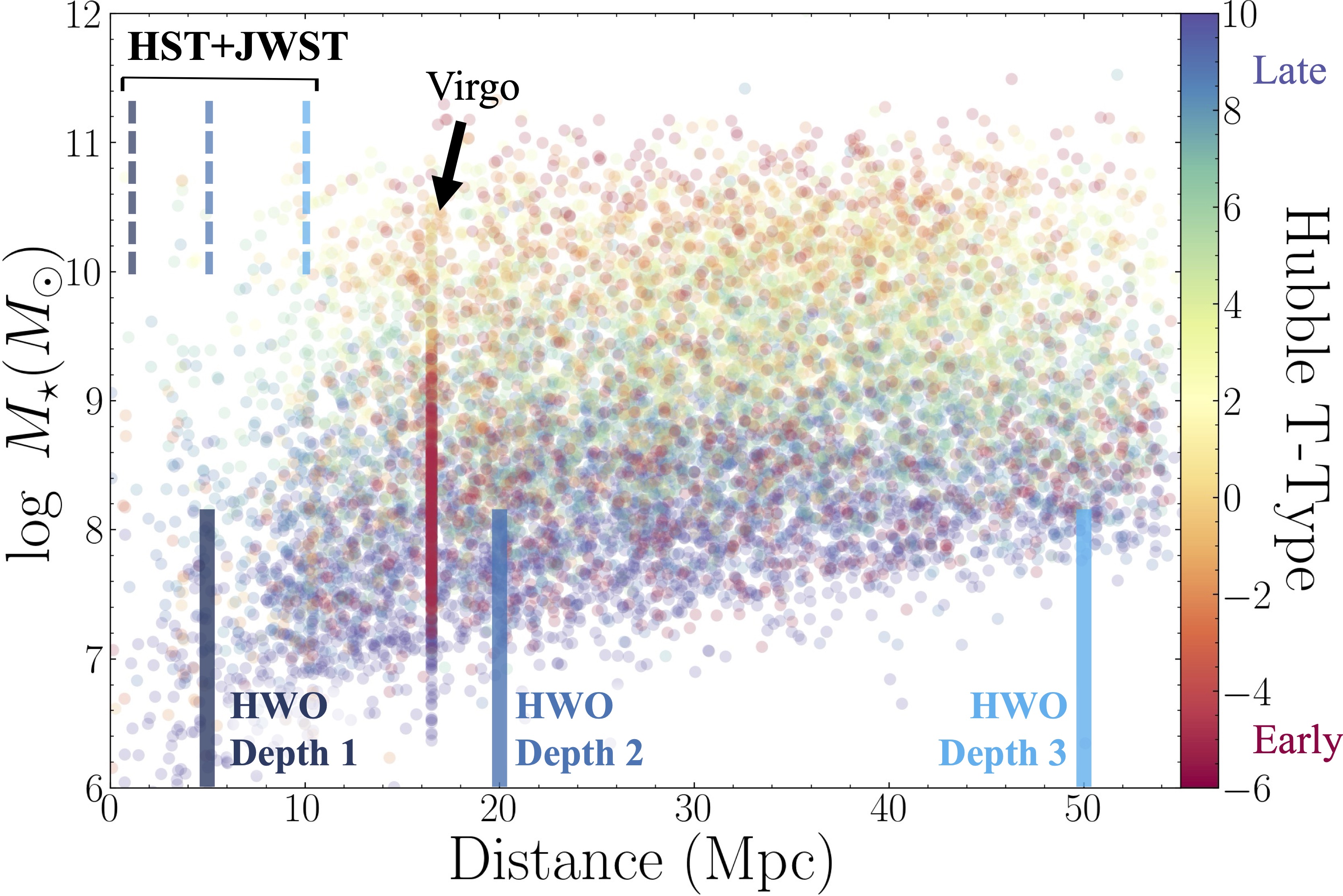}
    \caption{Distance vs.\ Stellar Mass for over 15,000 known galaxies out to 50\,Mpc \citep{Ohlson2024}. Galaxies are color-coded by their Hubble ``T'' Type, with early type galaxies $T$\,$<$\,0 and late-type galaxies $T$\,$>$\,0. The three stellar luminosity depth limits -- RGB, sub-Red Clump, and sub-oMSTO -- are shown as lines, corresponding to the distances where they are achievable for massive galaxies, and color-coded according to Fig. \ref{fig:cmd-depth}. The distances for these 3 depths achievable with HST and JWST are shown as dashed lines, while those achievable with HWO are shown as solid lines. The increase in galaxy diversity accessible from JWST$\rightarrow$HWO is striking. The color bimodality of galaxies \citep[e.g.,][]{Baldry2004} is visible within HWO’s reach, as is the Virgo Cluster which is notable as a vertical line of enhanced density.}
    \label{fig:gal-pops}
\end{figure*}

HST unlocked the ability to study external galaxies in comparable detail to the Milky Way. JWST has improved on that ability, with slightly higher resolution and significantly greater sensitivity providing access to the resolved stars in more distant galaxies. Relative to the diverse galaxy populations found in the wider universe, this has been equivalent to wading further into the shallows from shore. HWO will truly represent a paradigm shift -- pushing the highly-resolved study of galaxies into the vast waters of the cosmic ocean. 

As laid out in \S\,\ref{sec:objectives}, with the right instrumental setup HWO will impact this science most significantly in three regimes of cosmic distance, which correspond to the luminosity of important regions of galaxy color--magnitude diagrams. In Fig. \ref{fig:gal-pops} we show the known population of massive galaxies out to 50 Mpc \citep{Ohlson2024}, annotated by the distances to which these 3 regimes of stellar luminosity are reachable for current facilities (HST+JWST), and for HWO. The increase in galaxy diversity is striking, with HWO bringing resolved-star studies to a population of galaxies representative of the full morphological diversity present in the present-day universe.

The observables for this science will be the fluxes (magnitudes) of stellar point sources at multiple wavelengths. We describe the details of the three CMD/stellar luminosity regions below, and summarize them in Table \ref{tab:params}. See Fig. \ref{fig:cmd-depth} for reference. \\

\input{tab1}

\noindent\textbf{\underline{Depth Level 1} -- Nearby Galaxies ($<$\,5 Mpc):} \textit{Individual Stars to the oldest MSTO \& Resolved Young Stellar Clusters}

\begin{itemize}[topsep=-5pt,rightmargin=0pt,leftmargin=12pt,itemsep=0pt]
\item Individual resolved stars down to the oldest Main Sequence Turnoff (oMSTO), ~1 Msun, can be resolved, providing precise CMD-based SFHs comparable to galaxies in the Local Group
\item Large populations of individual young star clusters can be resolved, providing direct measurement of the stellar IMF in hundreds of galaxies, across stellar mass, metallicity, and SFR \\
\end{itemize}

\noindent\textbf{\underline{Depth Level 2} -- Local Volume (5--20 Mpc):} \textit{Individual Stars to below the Red Clump}

\begin{itemize}[topsep=-5pt,rightmargin=0pt,leftmargin=12pt,itemsep=0pt]
\item Individual resolved stars down to Red Clump / Horizontal Branch can be resolved, providing precise CMD-based SFHs within the past 6\,Gyr
\item Maps of metallicity and spatially-mapped metal enrichment histories over the past 6\,Gyr 
\item Ability to connect galaxy structures and recent merger history to SFH for dozens of galaxies, including those in large cosmological structures such as the Virgo Cluster \\
\end{itemize}

\noindent\textbf{\underline{Depth Level 3} -- Low Redshift (20--50+ Mpc):} \textit{Resolving the Bulk Structure and Luminous Stars in Galaxies}

\begin{itemize}[topsep=-5pt,rightmargin=0pt,leftmargin=12pt,itemsep=0pt]
\item Individual luminous stars -- O/B/A Main Sequence stars, Blue- and Red-Supergiants, AGB and RGB
\item Maps of galaxy structure by bulk stellar populations, allowing precise measurement of spiral arms, bars, bulges, and warps
\item High-resolution maps of interstellar dust and reddening 
\item Tip of the RGB (TRGB) distances to hundreds of thousands of galaxies, providing precise 3D structure of the Universe out to 50 Mpc
\end{itemize}

\section{Description of Observations}
\label{sec:description}

\begin{figure*}[ht!]
    \centering
    \includegraphics[width=0.7\linewidth]{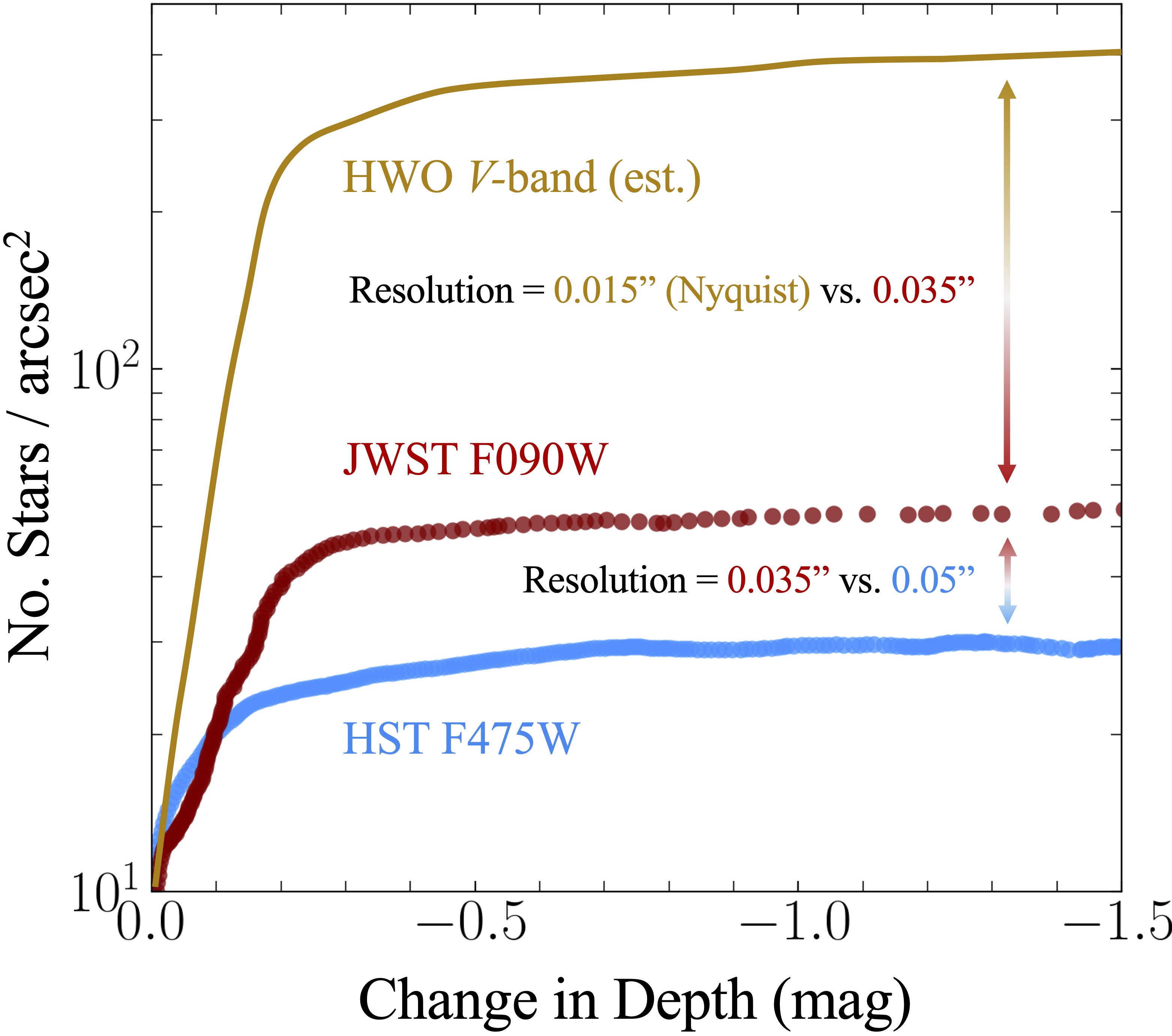}
    \caption{The impact of stellar crowding on limiting depth for different the different resolutions achieved by HST, JWST, and HWO (expected). Measurements from stars in M33 are shown for the HST UVIS g-band filter, F475W, and the F090W filter on JWST’s NIRCam. The PSF of both UVIS and NIRCam are sampled similarly at these respective wavelengths (and not particularly well). A relatively small improvement in the pixel scale from UVIS to NIRCam, however, results in an ability to resolve stars at densities nearly twice as high. With NIRCam, stars down to the RC can now be resolved in the disks of galaxies out to approximately 5\,Mpc. We also show the expectation for a $V$-band filter on HWO, with a resolution of 0\farcs015 (Nyquist sampled with 0\farcs01 pixels), following the LUVOIR design. Such a camera could resolve stars at on-sky densities nearly an order of magnitude higher than JWST, allowing observations of RC-depth stars in the main bodies of galaxies at distances up to 20 Mpc, and perhaps further. Individual luminous stars, such as RGBs, would be resolvable to 50+ Mpc. See Tables \ref{tab:params} \&\ \ref{tab:obs}.}
    \label{fig:crowding}
\end{figure*}

Here we describe in detail the technological requirements that HWO must meet to achieve the observational goals laid out in this science case. These requirements are collated in Table \ref{tab:obs}. 

\subsection{Well-Sampled, Diffraction-Limited Imaging Across the near-UV and Optical}

The primary hurdles to breakthrough science in the studying the resolved stars of nearby galaxies are resolution and sensitivity. A mirror comparable in size to JWST (i.e.\ 6.5\,m or greater) will provide a platform to meet these sensitivity and resolution requirements. However, realizing the full potential of this platform in the UV and optical will require a breakthrough in imaging quality for a space observatory, over a FOV similar to HST and JWST. 

As shown in Fig. \ref{fig:crowding}, JWST now represents the state-of-the-art in imaging resolution when compared to both HST at near-infrared wavelengths and in the UV/Optical wavelength range. When a field becomes `crowded', no additional stars can be resolved by the detector, and depth is no longer limited by S/N of an isolated point source, but by the number of stars in the population visible at that depth. The `crowding limit' can therefore be shallower than the nominal depth for deep observations. This crowding limit is reached when the density of stars as a function of depth hits a plateau --- the points at which the curves in Fig.\,\ref{fig:crowding} asymptote. JWST's improved resolution has resulted in a substantial improvement in the ability to measure stellar photometry in crowded fields -- boasting a nearly factor of 2 improvement over HST at UV and blue optical wavelengths. This gain is almost entirely due to JWST’s modestly better resolution and correspondingly smaller pixel scale. Combined with JWST’s improved sensitivity, this corresponds to an ability to measure stellar photometry in the main bodies of large galaxies roughly twice as far away as HST -- out to approximately 4--5\,Mpc. 

\input{tab2}

A similar-size mirror observing at UV/optical wavelengths, paired with modern detector technology, has the potential to provide a breakthrough in both resolution and depth, enabling our ability to measure crowded stellar photometry out to the edge of the Local Volume. As shown in Fig.\,\ref{fig:crowding}, this would correspond to an ability to resolve stellar point sources at on-sky densities nearly 100 times greater than achievable with JWST. The improved native resolution in the optical can be fully exploited by implementing detectors that are sufficiently well-sampled for much of the wavelength range of operation. The LUVOIR Final Report \citep{TheLUVOIRTeam2019} proposes a High Definition Imager, with 10 mas pixels and diffraction-limited imaging down to 550\,nm. This is exactly the level of improvement needed to achieve truly breakthrough science in resolved star studies of Local Volume galaxies (see Fig. \ref{fig:gal-pops}). 

\subsection{High-spatial Resolution IFU Spectroscopy}

Developing an ultra-high definition imager for HWO would naturally pair with a high-spatial resolution IFU instrument. As discussed above, resolved and semi-resolved spectroscopy of stars in galaxies provides invaluable information about their dynamical and chemical evolution \citep[e.g.,][]{Gibson2024}. HWO could be the first space observatory to allow ``crowded field spectroscopy'' \citep[e.g.,][]{Roth2019,Mainieri2024} in the main bodies of nearby galaxies. To meaningfully complement the imaging capabilities laid out in this case, this IFU would not only need to be high-spatial resolution, but would need to have a world-leading FOV of $>$1\arcmin.  

\acknowledgements
{\bf Acknowledgements:} AS is supported by NASA through the NASA Hubble Fellowship grant HST-HF2-51567 awarded by the Space Telescope Science Institute, which is operated by the Association of Universities for Research in Astronomy, Inc., for NASA, under contract NAS5-26555. TF acknowledges support from an appointment through the NASA Postdoctoral Program at the NASA Astrobiology Center, administered by Oak Ridge Associated Universities under contract with NASA.

\bibliography{author.bib}

\end{document}

%% file: tab1.tex
\begin{table*}[!ht]
\caption{Resolved Stars as a Function of Distance with HWO. The three depth tiers are color-coded as in Figs.\,\ref{fig:cmd-depth} \&\ \ref{fig:gal-pops}.}
\vspace{-8pt}
\begin{center}
{\small\label{tab:params}
\begin{tabular}{lllll} 
\tableline
\rowcolor{gray!15}
\textbf{Parameter} & \textbf{State of the Art} & \textbf{Incremental} & \textbf{Substantial Progress} & \textbf{Major Progress} \\
\rowcolor{gray!15}
(Distance) & (Current) & (Enhancing) & (Enabling) & (Breakthrough) \\
\tableline

\cellcolor{LightBlue!60}{
\parbox{4cm}{\raggedright \textbf{\underline{Depth 1}:} \textbf{13 Gyr SFHs + IMF} \\
Requires detecting stars to 1 mag below the oldest MSTO (oMSTO) + resolving individual young star clusters}} & 
\parbox{2cm}{Local Group \\ ($<$1 Mpc)} &
\parbox{2cm}{$\sim$1.5 Mpc \\ (near-field)} & 
\parbox{2.5cm}{\raggedright 2--2.5 Mpc \\ (Nearest isolated LMC-mass galaxies)} &
\parbox{4cm}{\smallskip\smallskip\raggedright \textbf{5 Mpc:} 
\begin{itemize}[itemsep=0pt,topsep=0pt,leftmargin=9pt]
    \item Dozens of large nearby galaxies w/ complete SFHs
    \item 1000s of star clusters across galaxy mass, metallicity, and SFR properties
\end{itemize}
\smallskip\smallskip} \\

\cellcolor{MedBlue!60}{
\parbox{4cm}{\raggedright \textbf{\underline{Depth 2}:} \textbf{6 Gyr SFHs} \\
Requires detecting individual stars to 1 mag below the Red Clump}} & 
\parbox{2cm}{Local Group \\ (M31 and M33)} &
\parbox{2cm}{\raggedright $\sim$3--4 Mpc \\ (Nearest MW-mass Groups)} & 
\parbox{2cm}{\raggedright $\sim$5--7 Mpc \\ (numerous large nearby galaxies)} &
\parbox{4.2cm}{\smallskip\smallskip\raggedright \textbf{20 Mpc:} 
\begin{itemize}[itemsep=0pt,topsep=0pt,leftmargin=9pt]
    \item 100s of galaxies w/ CMD-based SFHs (dozens of MW-analogs)
    \item Access to ongoing galaxy mergers, such as the Antennae
    \item First-ever access to true evolutionary histories of large-scale structures such as Virgo
\end{itemize}
\smallskip\smallskip} \\

\cellcolor{DarkBlue!60}{
\parbox{4cm}{\raggedright \textbf{\underline{Depth 3}:} \textbf{Galaxy Stellar Morphology + Dust Structure} \\
Requires resolving individual luminous Massive, AGB, and RGB stars}} & 
\parbox{2cm}{\raggedright $\sim$3--4 Mpc \\ (Nearest MW-mass Groups)} &
\parbox{2cm}{\raggedright $\sim$5--7 Mpc \\ (numerous large nearby galaxies)} & 
\parbox{2.5cm}{\raggedright $\sim$10--12 Mpc \\ (edge of current HST/JWST Volume)} &
\parbox{4cm}{\smallskip\smallskip\raggedright \textbf{50 Mpc:} 
\begin{itemize}[itemsep=0pt,topsep=0pt,leftmargin=9pt]
    \item 1000s of large galaxies with resolved luminous populations
    \item Resolved structure and characterization of spiral arms, warps, bars, dust, etc.
\end{itemize}
\smallskip\smallskip} \\

\tableline\
\end{tabular}
}
\end{center}
\end{table*}

%% file: tab2.tex
\begin{table*}[ht!]
\caption{\textbf{Observational Requirements} -- The pixel scales presented in the table below are calculated assuming the scaling shown in Fig. \ref{fig:crowding}. Listed magnitude limits assume exposure times of 10 hours, following the LUVOIR Final Report \citep{TheLUVOIRTeam2019}.}
\vspace{-8pt}
\begin{center}
{\small\label{tab:obs}
\begin{tabular}{lllll} 
\tableline
\rowcolor{gray!15}
\textbf{Parameter} & \textbf{State of the Art} & \textbf{Incremental} & \textbf{Substantial Progress} & \textbf{Major Progress} \\
\rowcolor{gray!15}
(Distance) & (Current) & (Enhancing) & (Enabling) & (Breakthrough) \\
\tableline
\noalign{\smallskip}

\parbox{2.5cm}{\raggedright Imaging of resolved stars in nearby massive galaxies} & 
\parbox{3cm}{\raggedright Highest available imaging resolution = JWST IR} &
\parbox{3cm}{\raggedright JWST-resolution imaging in UV/optical (0\farcs031 pixels vs. 0\farcs04 for HST)} & 
\parbox{3cm}{\raggedright 0\farcs02--0\farcs025 pixels in UV/optical (severely under-sampled)} &
\parbox{3.8cm}{\raggedright Well-sampled, diffraction-limited UV/optical/IR imaging} \\

\noalign{\smallskip}\noalign{\smallskip}\noalign{\smallskip}

\parbox{2.3cm}{\raggedright Wavelength Range} & 
\parbox{3cm}{NIR} &
\parbox{3cm}{UBVRI + NIR} & 
\parbox{3cm}{UBVRI + NIR} &
\parbox{3.8cm}{NUV + UBVRI + NIR} \\

\noalign{\smallskip}\noalign{\smallskip}\noalign{\smallskip}

\parbox{2.3cm}{Pixel Scale} & 
\parbox{3cm}{0\farcs031 at NIR} &
\parbox{3cm}{0\farcs031 at UV/optical} & 
\parbox{3cm}{0\farcs02 at UV/optical} &
\parbox{3.8cm}{\raggedright 0\farcs01 at UV/optical (Nyquist sampled at $V$)} \\

\noalign{\smallskip}\noalign{\smallskip}\noalign{\smallskip}

\parbox{2.3cm}{FOV} & 
\parbox{2.3cm}{\raggedright 
0.0031 deg$^2$\ \\ (HST ACS) \\

0.0027 deg$^2$\ \\ (JWST NIRCam) \\

0.002 deg$^2$\ \\ (HST UVIS)} &
\parbox{3cm}{\raggedright 0.0027 deg$^2$\ across UV/Optical/NIR \\ (NIRCam FOV)} & 
\parbox{3cm}{\raggedright 0.0031 deg$^2$\ across UV/Optical/NIR (ACS FOV)} &
\parbox{3.8cm}{\raggedright 0.007 deg$^2$\ across UV/Optical/NIR (increased FOV; 300\arcsec\ on a side) \\\vspace{5pt}

Equivalent to 23 kpc at 16 Mpc} \\

\noalign{\smallskip}\noalign{\smallskip}\noalign{\smallskip}

\parbox{2.5cm}{\raggedright Best Spatial Resolution} & 
\parbox{2.3cm}{\raggedright $\sim$0\farcs035 at 1\,\textmu m} &
\parbox{3cm}{\raggedright $\sim$0\farcs035 across UV/optical} & 
\parbox{3cm}{\raggedright $\sim$0\farcs025 at $V$} &
\parbox{3.8cm}{\raggedright Diffraction limited at 550\,nm, $\sim$0\farcs015 } \\

\noalign{\smallskip}\noalign{\smallskip}\noalign{\smallskip}

\parbox{2.3cm}{\raggedright Magnitude of target in chosen bandpass} & 
\parbox{2.3cm}{\raggedright F814W\,$<$\,29 \\ (HST) \\\vspace{5pt}

F200W\,$<$\,30 \\ (JWST)} &
\parbox{3cm}{\raggedright UV\,$<$\,29.5 \\ 
$B\,{<}$\,31 \\
$I\,{<}$\,30 \\
(Red Clump @ 10 Mpc)} & 
\parbox{3cm}{\raggedright UV\,$<$\,30.5 \\ 
$B\,{<}$\,32 \\
$I\,{<}$\,31 \\
(Red Clump @ 15 Mpc)} &
\parbox{3.8cm}{\raggedright UV\,$<$\,32.5 \\ 
$B\,{<}$\,34 \\
$I\,{<}$\,33 \\\vspace{5pt}
oMSTO+1 mag @ 5 Mpc \\\vspace{5pt}
Red Clump + 1 mag @ 20 Mpc \\\vspace{5pt}
TRGB+2 mag @ 50 Mpc} \\

\noalign{\smallskip}
\tableline\
\end{tabular}
}
\end{center}
\end{table*}